\documentclass{article}
\usepackage[utf8]{inputenc}
\usepackage{enumitem}
\usepackage{amsmath}
\begin{document}

  \begin{center}

 \Large Fields and strings on non commutative q-deformed spaces

 \end{center}

 \begin{center}
     Poula Tadros\\

     Department of Applied Physics, Aalto University School of Science, FI-00076 Aalto, Finland.\\

     email:poulatadros9@gmail.com

\end{center}

\abstract{We study scalar field and string theory on non commutative q-deformed spaces. We define a product of functions on a non commutative algebra of functions resulting from the q-deformation analog to the Moyal product for canonically non commutative spaces. We then give the general procedure to define scalar field and classical string theories on the mentioned spaces, we argue that the resulting theories will have enlarged sets of both spacetime and internal  symmetries which can be used to study gravitational effects due to the q-deformation.}


\section{Introduction}

Non commutative geometry was introduced in string theory in [1] where it was shown that the coordinates of the endpoints of strings on D-branes in presence of Neveu-Schwartz field is non commutative. In field theory it was even older where Yang-Mills theory on non commutative torus was introduced [2].\\

The main motivation to introduce non commutative space times is field theory is explained in [3,4]. In quantum mechanics Heisenberg uncertainty principle states that at small distance scales there is a large uncertainty in momentum measurement i.e. energy can reach very high values in small space distance (close to the Planck scale), but according to the general theory of relativity, high energy in sufficiently small distance scale creates a black hole preventing measurement of position to be fully certain i.e. there is uncertainty in position measurement in small scales, this can only be achieved by introducing non commutativity in space time. Notice that this implies non locality in the theory.\\

Since the introduction of non commutativity in field and string theories a lot of progress has been made in all directions including classical and quantum field theories, theories of gravity and string theory. However, the non commutativity used is the canonical non commutativity which does not capture the mathematical structure of the given field or string theory and it is clear that it was imposed by hand. In this article we use another type of non commutativity, the q-deformation, to study classical scalar field theory and the consequences on string theory. In section 2, we review the most popular types of non commutativity on space times and motivate the choice of q-deformation as the non commutativity of choice. In section 3, we define a product of functions on q-deformed spaces similar to the Moyal product on canonically non commutative spaces and show the procedure failed for Lie-type non commutativity. In section 4, we study scalar field theory on q-deformed space time. In section 5, we study string theory on the same space time. In section 6, we discuss the symmetries of the non commutative theories, we show that there are more symmetries in the non commutative theories than the corresponding commutative ones, then use this to argue that we can define theories with dynamical spacetimes by quantizing the spacetime symmetry group of the theory. \\

\section{Types of non commutativity}
This section is dedicated to review three types of non commutativity of space times

\subsection{Canonical non commutativity}

It is the simplest type and is the one used in physics literature, it was introduced in [5], it is defined by imposing the following commutation relations on the space time
$$[x^{\mu},x^{\nu}]=i\theta^{\mu \nu},$$
where $x^{\mu}$ are the space time coordinates and $\theta^{\mu \nu}$ is a constant, anti symmetric matrix.\\

Canonical non commutativity corresponds to smearing of the space time, it can be easily seen from solutions of polynomial equations on the commutative space compared to its non commutative counterpart.\\

As an example consider the two dimensional Euclidean space with coordinates $x$ and $y$ with the commutation relation $[x,y]=k$, where $k$ is a positive constant, and consider the polynomial equation $(x-y)^2=0$.\\
In the usual commutative space, the solution to the above equation is $x=y$ which is a straight line with slope = 1 and passing through the origin. However, in the corresponding non commutative space the equation can be written as $x^2-2yx+y^2=k$ whose solutions are two parallel straight lines separated by a distance proportional to $k$, when $k=0$ the two straight lines coincide and we recover the solution on the commutative space.\\

Note that this procedure is valid regardless of whether or not there are additional mathematical structures on the space, the smearing is carried out the same way. That is why we need more complicated non commutativity to use in physics.

\subsection{Lie-type non commutativity}

In this case the coordinates has a Lie algebra structure i.e. the commutation relations can capture a Lie algebra structures if defined on the space time for example like in field theories [6]. The commutation relations are given by

$$[x^{\mu}, x^{\nu}]=if^{\mu \nu}_{\rho}x^{\rho},$$
where $f^{\mu \nu}_{\rho}$ are the structure constants of the defined Lie algebra. However, this type is not useful because Lie structures are rigid i.e. any small deformation of a Lie algebra is isomorphic to the Lie algebra. This leads to difficulties in defining products of functions on the resulting non commutative space as we will see in the next section.

\subsection{q-deformations}

This type was introduced to solve the rigidity problem for Lie algebras. The main idea is to replace Lie group with a flexible structure which is called quantum groups, for more details on the theory of quantum groups see [7,8].\\

The commutation relations are given by

$$x^{\mu} x^{\nu}= \frac{1}{q} R^{\mu \nu}_{\sigma \tau}x^{\sigma}x^{\tau},$$
where $q$ is a parameter and $R^{\mu \nu}_{\sigma \tau}$ is the R-matrix of the quantum group defined on the space.\\

In this space a Lie algebra is replaced by a non commutative Hopf algebra with deformation parameter q. Hopf algebras are considered deformations of the universal enveloping algebra of the Lie group. The resulting space is deformed according to the Lie group on the space and on the parameter q, this is the simplest way to deform a space time while capturing the full structure of the space, other more complicated approaches can be studied such as deforming with more than one parameter [9,10] but it is beyond the scope of the article.

\section{Moyal-like product on q-deformed spaces}
Here, we define a non commutative product of functions on q-deformed spaces i.e. non commutative spaces in the R-matrix formalism.

\subsection{Moyal product on canonically non commutative spaces}

We begin with reviewing the original Moyal product. On Canonically non commutative spaces, the algebra of functions is replaced by a non commutative $C*$ algebra, the Moyal product is the product of functions on the non commutative algebra. Its formula can be derived as follows:\\
Consider two functions $f(x)$ and $g(x)$, their Fourier transforms are
$$f(x)=\int \frac{d^Dk}{(2\pi)^D}\bar{f(k)}e^{ik_ix^i},$$
$$g(x)=\int \frac{d^Dk'}{(2\pi)^D}\bar{g(k)}e^{ik_jx^j},$$
The product on the non commutative space is
$$f(x)\star g(x)= \int \frac{d^Dk d^Dk'}{(2\pi)^{2D}} \bar{f(k)}\bar{g(k)}e^{ik_ix^i}e^{ik_jx^j}.$$
Using Baker-Campbell-Hausdorff formula we get

$$f(x)\star g(x)= \int \frac{d^Dk d^Dk'}{(2\pi)^{2D}} \bar{f(k)}\bar{g(k)}e^{ik_ix^i}e^{ik_jx^j}e^{i/2k_ik_j\theta^{ij}}$$
$$=\int \frac{d^Dk d^Dk'}{(2\pi)^{2D}} \bar{f(k)}\bar{g(k)}e^{ik_ix^i}e^{ik_jx^j}(1+\sum_{n=1}^{\infty}(\frac{i}{2})^n \frac{1}{n!}(k_ik_j\theta^{ij})^n)$$
$$=f(x)g(x)+\sum_{n=1}^{\infty}(\frac{i}{2})^n \frac{1}{n!} \theta^{i_1j_1}...\theta^{i_nj_n}\partial_{i_1}\partial_{i_2}...\partial_{i_n}f\partial_{j_1}...\partial_{j_n}g.$$

\subsection{Product of functions on q-deformed space}

We follow the same procedure to define a product on q-deformed spaces. The non commutativity is given by 
$$x^{\mu}x^{\nu}=\frac{1}{q}R_{\sigma \tau}^{\mu \nu}x^{\sigma}x^{\tau}.$$
It can be written as a commutation relation as
$$[x^{\mu},x^{\nu}]=Q_{\sigma \tau}^{\mu \nu}x^{\sigma}x^{\tau},$$
where $Q_{\sigma \tau}^{\mu \nu}=\frac{1}{q}R_{\sigma \tau}^{\mu \nu}-\delta_{\tau}^{\mu} \delta_{\sigma}^{\nu}.$ Note that at $q=1$ we have $Q_{\sigma \tau}^{\mu \nu}=0$ and we recover commutativity.\\

Now again two functions $f(x)$ and $g(x)$. the product of their Fourier transforms is
$$f(x)\star g(x)= \int \frac{d^Dk d^Dk'}{(2\pi)^{2D}} \bar{f(k)}\bar{g(k)}e^{ik_ix^i}e^{ik_jx^j},$$

and using Baker-Campbell-Hausdorff formula we get

$$f(x)\star g(x)= \int \frac{d^Dk d^Dk'}{(2\pi)^{2D}} \bar{f(k)}\bar{g(k)}e^{ik_ix^i}e^{ik_jx^j}$$ $$e^{x^i+x^j+1/2[x^i,x^j]+1/12[x^i,[x^i,x^j]]-1/12[x^j,[x^i,x^j]]+...},$$

after some calculations we have
$$[x^i,[x^i,x^j]]=(Q^{ij}_{nl}Q_{ab}^{il}+Q_{cb}^{ij}Q_{na}^{im})x^nx^ax^b,$$
$$[x^i,[x^i,x^j]]=(Q^{ij}_{nl}Q_{ab}^{jl}+Q_{cb}^{ij}Q_{na}^{jm})x^nx^ax^b,$$

Substituting we get

$$f(x)\star g(x)= \int \frac{d^Dk d^Dk'}{(2\pi)^{2D}} \bar{f(k)}\bar{g(k)}e^{ik_ix^i}e^{ik_jx^j}$$
$$e^{x^ik_i+x^jk'_j+1/2 Q^{ij}_{mn}x^mx^nk_ik'_j+1/12k_ik'_j(Q^{ij}_{nl}Q_{ab}^{il}-Q_{cb}^{ij}Q_{na}^{im}-Q^{ij}_{nl}Q_{ab}^{jl}+Q_{cb}^{ij}Q_{na}^{jm})x^nx^ax^b+...}.$$

In string theory it is reasonable to assume that the parameter $q$ is close to 1 since the operator $Q$ is related to the string length which is assumed to be very small. In field theory the assumption is reasonable as well since in this case $Q$ is related to the area of the space time where general relativity breaks i.e. quantum gravity scale which is assumed to be very small. Thus, the series converges and all the exponentials are well defined. If we ignore the higher orders in the last exponent we get a formula similar to the Moyal product:

$$f(x)\star g(x)= \int \frac{d^Dk d^Dk'}{(2\pi)^{2D}} \bar{f(k)}\bar{g(k')}e^{ik_ix^i}e^{ik'_jx^j}e^{1/2 Q^{ij}_{mn}x^mx^nk_ik'_j}$$
$$=\int \frac{d^Dk d^Dk'}{(2\pi)^{2D}} \bar{f(k)}\bar{g(k)}e^{ik_ix^i}e^{ik'_jx^j}(1+\sum_{n=1}^{\infty}(1/2)^n\frac{1}{n!} (Q^{ij}_{mn}x^mx^nk_ik'_j)^n)$$
$$=f(x)g(x)+\sum_{p=1}^{\infty}(1/2)^p\frac{1}{p!} x^{m_1}x^{l_1}Q_{m_1l_1}^{i_1j_1}...x^{m_p}x^{l_p}Q_{m_pl_p}^{i_pj_p}\partial_{i_1}\partial_{i_2}...\partial_{i_p}f(x) \partial_{j_1}...\partial_{j_p}g(x).$$
The formula captures the mathematical structures on the space time in the form of the $Q$ operator and the deformation is subsequently transformed into these structures leading to additional symmetries at least in string theory.\\

From this procedure we can see that Lie-type non commutativity presents difficulties in defining the product since the product would be of the form 
$$f(x)\star g(x)= \int \frac{d^Dk d^Dk'}{(2\pi)^{2D}} \bar{f(k)}\bar{g(k)}e^{ik_ix^i}e^{ik_jx^j}$$
$$e^{1/2(f^{ij}_mx^m k_ik'_j)+1/12 k_ik'_j(f^{ij}_{nl}f_{ab}^{il}-f_{cb}^{ij}f_{na}^{im})+...}.$$
The series in the exponential generally diverges and the product can not be defined.

\section{Scalar field theory on q-deformed space}

In this section we study massive scalar field theory on a flat non dynamical q-deformed non commutative space time. While there are attempts to define field theories on deformed spaces, the used non commutativity is usually the canonical type or the use of quantum groups is limited to defining differential structure on a specific examples of spaces [11-14], conformal field theory also was studied on deformed spaces [15], however the deformations considered are introduced by hand and does not introduce non commutativity i.e. the deformed manifold is another manifold with no additional structure.\\

The Lagrangian of the theory is 
$$\mathcal{L}=\partial_{\mu}\phi \partial^{\mu} \phi -m^2 \phi^2,$$
where $\phi$ is the scalar field which we will assume is infinitely differentiable, and $m$ is the mass.\\

Now we perform the deformation quantization:\\
The first step is to perform the q-quantization of the symmetry group in this case $U(1)$. To do this we write its universal enveloping algebra which is a $C*$ algebra of functions generated by the function $z \in \mathcal{U}(U(1)) \rightarrow e^{(iz)} \in \mathbf{C}$. We notice that the algebra is commutative then its deformations are equivalent to itself, i.e. no contribution from the symmetry group to non commutativity and the product on the non commutative space is equivalent to the product on the original manifold.\\

The second step is to replace the manifold on which the field theory is defined with a non commutative locally compact topological space. On this manifold the derivatives are q-deformed into Jackson derivatives.\\

The new q-deformed Lagrangian will be

$$\mathcal{L}_q=D_{q\mu}\phi D_q^{\mu} \phi -m^2 \phi^2,$$

Now we relate the theory on the non commutative topological space to the theory on the commutative manifold (i.e. transforming the non commutative theory back to the commutative manifold) using the formula
 $$D_{q\mu}(f(x))= \partial_{\mu}f+ \sum_{k=1}^{\infty}\frac{(q-1)^k}{(k+1)!}x_{\mu}^k f^{(k+1)}(x) ,$$ where $f^{(k)}$ is the k th ordinary derivative of f.\\

The resulting Lagrangian on the commutative manifold is
$$\mathcal{L}_q=\partial_{\mu}\phi \partial^{\mu} \phi -m^2 \phi^2 + 2\partial_{\mu}\phi \sum_{k=1}^{\infty}\frac{(q-1)^k}{(k+1)!}x^{\mu k} \phi^{(k+1)} $$
$$+ \sum_{l,m=1}^{\infty}\frac{(q-1)^{(l+m)}}{(m+1)!(l+1)!}\phi^{(l+1)}x_{\mu}^k x^{\mu l} \phi^{(m+1)}.$$

The first two terms of the Lagrangian is the non commutative original theory and the rest are the contributions of non commutativity from replacing the non commutative topological space with the original commutative manifold. The theory is q-deformed i.e. if $q=1$ then we recover the original theory.\\

The additional terms are non local as expected and contain an infinite series of higher (ordinary) derivatives of the field $\phi$. \\

\section{String theory on q-deformed space}

String theory follows the same q-quantization procedure as field theory but with richer geometry since the fundamental object is one dimensional. Here we establish the connection between the $Q$ operator defined above and the length of the string, then give the general procedure of defining a string theory on q-deformed space.\\

The uncertainty in position in case of q-deformed spaces can be calculated to be
$$\Delta x^i \Delta x^j \geq \frac{1}{2}<x^{\mu}Q_{\mu \nu}^{ij}x^{\nu}>,$$
where $<x^{\mu}Q_{\mu \nu}^{ij}x^{\nu}>$ is the expectation value of the quadratic form of the operator $Q$.\\

Following the same argument as [2], we find that the length of the string squared is proportional to the above expectation value
$$<x^{\mu}Q_{\mu \nu}^{ij}x^{\nu}> \ \ \propto l_s^2.$$

This implies that the string's length depends on the geometry of the non commutative space i.e. depends on the string theory in question, and is determined by the R-matrix of the quantized group.\\

The procedure on a static spacetime is as follows:\\
\begin{enumerate}
    \item Determine the symmetry group of the theory and find the corresponding quantum group.
    \item Use the product presented in section 3 instead of the usual product and Jackson's derivative instead of the usual derivative.
    \item Use the corresponding formulae to relate back to the original manifold as we did in section 4, this usually leads to infinite series of higher derivatives in the Lagrangian.
\end{enumerate}

\section{Symmetries and theories on dynamical spacetimes}

The first step of q-quantization is to replace the symmetry group with a quantum group which is a deformation of its universal enveloping algebra. This gives more symmetries than the commutative theory by definition.\\

In field and string theory, symmetries are classified into spacetime symmetries and internal symmetries, spacetime symmetries relates directly to the ambient manifold on which the field/string theory is defined while internal symmetries are additional structure on the manifold. While on static spacetime (disregarding gravity) only the internal symmetry group is to be q-deformed, the spacetime symmetry group must contribute to the R-matrix if dynamic spacetimes are to be studied, the deformations of the spacetime symmetry should lead to effects on the gravitational aspects of the theory like changes in curvature, singularities, etc. Similar studies of non commutativity's effects on gravity are found in [] but uses the canonical non commutativity, using q-deformations to study gravity is a subject of future research.

\section{Conclusion and outlook}

The results presented in this paper showed that a product of functions on a q-deformed space at least for small deformations exists and is well defined, we give an explicit formula in the paper. We also showed that field and string theory can be defined on q-deformed manifolds but having enlarged set of symmetries and extra features depending on the theory and the manifold in question.\\

A possible direction of future research is to study the enlarged set of symmetries due to q-deformations as well as their mathematical and the phenomenological implications. Another direction is to study more complicated field/string theories and find ways to define higher spin fields on such spaces.

\section*{Acknowledgments}
We would like to thank Dr.Ivan Kolar for the useful discussions on the topic.


\section*{References}

\begin{enumerate}[label={[\arabic*]}]

    \item Seiberg, N. and Witten, E. (1999) “String theory and noncommutative geometry,” Journal of High Energy Physics, 1999(09), pp. 032–032.

    \item Szabo, R. (2003) “Quantum field theory on noncommutative spaces,” Physics Reports, 378(4), pp. 207–299.

    \item Doplicher, S., Fredenhagen, K. and Roberts, J.E. (1995) “The quantum structure of spacetime at the Planck scale and Quantum Fields,” Communications in Mathematical Physics, 172(1), pp. 187–220.  

    \item Ahluwalia, D.V. (1994) “Quantum measurement, gravitation, and locality,” Physics Letters B, 339(4), pp. 301–303.

    \item C. S. Chu and P. M. Ho, Noncommutative open string and D-brane, Nucl. Phys.
B 550, 151 (1999) [hep-th/9812219].

\item  B. Jurco, S. Schraml, P. Schupp and J. Wess, Enveloping algebra valued gauge
transformations for non-Abelian gauge groups on non-commutative spaces, Eur.
Phys. J. C17, 521 (2000) [hep-th/0006246].

\item Chaichian, M. and Demichev, A.P.  Introduction to quantum groups. Singapore: World Scientific (1996). 

\item A. Klimyk and K. Schmudgen, Quantum Groups and Their Representations,
Springer (1997).

\item Hu, N.H. and Pei, Y.F. (2008) “Notes on 2-parameter Quantum Groups I,” Science in China Series A: Mathematics, 51(6), pp. 1101–1110.

\item Hu, N. and Pei, Y. (2012) “Notes on two-parameter quantum groups, (II),” Communications in Algebra, 40(9), pp. 3202–3220. 

\item Wulkenhaar, R. (2006) “Field theories on deformed spaces,” Journal of Geometry and Physics, 56(1), pp. 108–141. 

\item Grosse, H., Madore, J. and Steinacker, H. (2001) “Field theory on the Q-deformed fuzzy sphere I,” Journal of Geometry and Physics, 38(3-4), pp. 308–342.

\item Grosse, H., Madore, J. and Steinacker, H. (2002) “Field theory on the Q-deformed Fuzzy Sphere II: Quantization,” Journal of Geometry and Physics, 43(2-3), pp. 205–240. 

\item BARDEK, V., DOREŠIĆ, M. and MELJANAC, S. (1994) “An example of Q-deformed field theory,” International Journal of Modern Physics A, 09(23), pp. 4185–4194.

\item Minahan, J., Naseer, U. and Thull, C. (2021) “Conformal field theories on deformed spheres, anomalies, and  supersymmetry,” SciPost Physics, 10(3).

\end{enumerate}

\end{document}